\documentclass[journal=jacsat,manuscript=article]{achemso}

\usepackage[version=3]{mhchem} \usepackage{soul}
\usepackage{xcolor}
\usepackage{setspace}
\usepackage{hyperref}

\hypersetup{
    colorlinks=true,
    linkcolor=blue,
    filecolor=magenta,      
    urlcolor=cyan,
    pdftitle={Overleaf Example},
    pdfpagemode=FullScreen,
    }

\makeatletter
\let\l@addto@macro\relax
\makeatother
\usepackage[fontsize=10pt]{scrextend}

\author{Alessio Stefani}
\affiliation[Institute of Photonics and Optical Science]{Institute of Photonics and Optical Science (IPOS), School of Physics, The University of Sydney, NSW 2006, Australia}
\author{Boris T. Kuhlmey}
\affiliation[Institute of Photonics and Optical Science]{Institute of Photonics and Optical Science (IPOS), School of Physics, The University of Sydney, NSW 2006, Australia}
\author{Justin Digweed}
\affiliation[The University of Sydney Nano Institute]{The University of Sydney Nano Institute (Sydney Nano), The University of Sydney, NSW 2006, Australia}
\author{Benjamin Davies}
\affiliation[Biomaterials and Tissue Engineering Research Unit]{Biomaterials and Tissue Engineering Research Unit, School of Biomedical Engineering, The University of Sydney, NSW 2006, Australia}
\author{Hala Zreiqat}
\affiliation[Biomaterials and Tissue Engineering Research Unit]{Biomaterials and Tissue Engineering Research Unit, School of Biomedical Engineering, The University of Sydney, NSW 2006, Australia}
\alsoaffiliation[Australian Research Council Training Centre for Innovative Bioengineering]
{Australian Research Council Training Centre for Innovative Bioengineering, Sydney, NSW 2006, Australia}
\author{Mohammad Mirkhalaf}
\affiliation[Biomaterials and Tissue Engineering Research Unit]{Biomaterials and Tissue Engineering Research Unit, School of Biomedical Engineering, The University of Sydney, NSW 2006, Australia}
\alsoaffiliation[Australian Research Council Training Centre for Innovative Bioengineering]
{Australian Research Council Training Centre for Innovative Bioengineering, Sydney, NSW 2006, Australia}
\alsoaffiliation[Australian Research Council Training Centre for Innovative Bioengineering]
{School of Mechanical, Medical and Process Engineering, Queensland University of Technology, 2 George St Brisbane, QLD 4000 Australia}
\author{Alessandro Tuniz}
\affiliation[Institute of Photonics and Optical Science]{Institute of Photonics and Optical Science (IPOS), School of Physics, The University of Sydney, NSW 2006, Australia}
\alsoaffiliation[The University of Sydney Nano Institute]
{The University of Sydney Nano Institute (Sydney Nano), The University of Sydney, NSW 2006, Australia}
\email{alessandro.tuniz@sydney.edu.au}

\title{Flexible terahertz photonic light-cage modules \\
for in-core sensing and high temperature applications}

\keywords{Terahertz photonics, hollow core waveguide, 3D printing, terahertz spectroscopy, antiresonant guidance}

\begin{document}


\newpage

\begin{abstract}
{Terahertz technology is a growing and multi-disciplinary research field, particularly in applications relating to sensing and telecommunications. A number of terahertz waveguides have emerged over the past few years, which are set to complement the capabilities of existing and bulky free space setups. In most terahertz waveguide designs however, the guiding region is physically separated from the surroundings, making any interaction between the guided light and the environment inefficient. 
Here we present photonic terahertz light cages (THzLCs) operating at terahertz frequencies, consisting of free-standing dielectric strands, which guide light within a central hollow core with immediate access to the environment. We experimentally show the versatility and design flexibility of this concept, by 3D-printing several cm-length-scale modules on the basis of a single design, using four different polymer- and ceramic- materials, which are either rigid, flexible, or resistant to high temperatures. We characterize both propagation- and bend- losses for straight- and curved- waveguides, which are of order $\sim$1\,dB/cm in the former, and $\sim$2-8\,dB/cm in the latter for bend radii below 10\,cm, and largely independent of the chosen material. Our transmission experiments are complemented by near-field measurements at the waveguide output, which reveal the antiresonant guidance for straight THzLCs, and a deformed fundamental mode in the bent waveguides, in agreement with numerical conformal mapping simulation models. We show that these THzLCs can be used either as: (i) flexible, reconfigurable, and bendable modular assemblies; (ii) in-core sensors of resonant structures contained \emph{directly inside} the hollow core; (iii) high-temperature waveguide sensors, with potential applications in industrial monitoring and sensing. The 3D-printed light cages presented are a novel and useful addition to the growing library of terahertz waveguides, marrying the waveguide-like advantages of reconfigurable, diffractionless propagation, with the free-space-like immediacy of direct exposure to the surrounding environment.}
\end{abstract}

\newpage

\section{Introduction}

Recent years have been marked by a rapid development in sources, detectors, and waveguides operating in the terahertz frequency range ($\sim 0.1-10\,{\rm THz}$), which has the unique capability of serving a number of far-reaching and diverse applications areas~\cite{jepsen2011terahertz, dhillon20172017}. These include sensing and spectroscopy (e.g., of gases~\cite{mittleman1998gas}, molecules~\cite{seo2020terahertz}, or  DNA~\cite{fischer2002far,ahmadivand2020terahertz}); imaging and security~\cite{liu2007terahertz, tuniz2013metamaterial,Atakaramians2017}, pharmaceutical research~\cite{taday2004applications}; broadband wireless communication (6G)~\cite{saad2019vision}; and industrial applications~\cite{tao2020non}. In spite of its significance, and even as the number of THz sources and detectors continue to expand~\cite{lewis2014review}, devices which operate at terahertz frequencies have yet to reach a level of maturity and sophisitication that is comparable with their photonic counterparts~\cite{shastri2021photonics}.

As a result, much effort has been dedicated to developing novel terahertz waveguides, particularly dielectric-based fiber-like devices that can guide and route terahertz radiation at will. These include solid-core step index waveguides~\cite{nielsen2009bendable}, sub-wavelength core fibers~\cite{roze2011suspended}, hollow-core fibers~\cite{lu2008terahertz, chen2010novel, setti2013flexible,bao2015dielectric, li2016flexible, Stefani:21}, metal tube waveguides~\cite{lu2010bending}, porous fibers~\cite{dupuis2009fabrication,atakaramians2009thz}, and Bragg fibers~\cite{skorobogatiy2007ferroelectric}, to name a few. More broadly, planar THz waveguides have also recently emerged, including photonic crystal-~\cite{tsuruda2015extremely}, step index-~\cite{nallappan2021terahertz} and topological~\cite{yang2020terahertz} waveguides, as well as modular circuit elements~\cite{cao2020additive}. Each structure has various advantages and disadvantages; we refer the reader to Refs.~\cite{humbert2019optical, islam2020terahertz, nallappan2021terahertz} for recent reviews on terahertz waveguides.

In index-guided waveguides, the environment can be weakly probed via the modes' evanescent fields; in hollow core waveguides, it is necessary to wait for any analytes -- be they gases~\cite{ritari2004gas}, liquids~\cite{wu2009ultrasensitive}, or scatterers~\cite{faez2015fast} -- to diffuse inside the walled-off area where the light is guided. Alternatively, one can use the light guided by the fiber core for remote endoscopy, but only at the waveguide output~\cite{lu2014terahertz}. These shortcomings can be addressed, for example, by introducing occasional hollow channels in the cladding that accelerate diffusion of gases from the environment into the core~\cite{hoo2010fast}, but such processes require significant post-processing. We note that single metal wires can guide terahertz radiation via radially polarized surface modes and with relatively low loss and in direct contact with the surrounding environment, but the associated coupling efficiency and confinement characteristics are notoriously poor~\cite{wang2004metal}. Multi-wire waveguides~\cite{mbonye2009terahertz} offer an attractive alternative, but typically rely on additional support to maintain the desired spacing and configuration~\cite{mbonye2009terahertz, dong2022versatile}, and are often enclosed entirely~\cite{tuniz2013metamaterial, cao2020additive, li2016flexible}.

In recent years, an alternative novel hollow core waveguide has emerged: the so-called ``photonic light cage''~\cite{jain2018hollow, jang2019light}. This geometry consists of a single ring of free-standing cylindrical dielectric strands of wavelength-scale diameter, arranged around a hollow core within which the light is guided. The appeal of this kind of geometry lies in its unique capability of providing direct access to the core via the spaces between the strands.
This effectively provides diffractionless propagation through free space, albeit at the cost of small propagation losses due to the leaky nature of the guided modes of such structures. 
This kind of waveguide is particularly attractive for applications which demand long-length light-matter interactions~\cite{davidson2021coherent}, e.g., gas~\cite{jain2018hollow} and liquid~\cite{kim2020optofluidic} sensing,  immediately containing any analytes within the waveguide core section. So far, such devices have been designed to operate in the visible and near-infrared regime~\cite{jain2018hollow, jang2019light, kim2020optofluidic, davidson2021coherent} (all fabricated via 3D-nanoprinting), only in straight configurations, with the transmittance and modes measured in the far field. To the best of our knowledge, a detailed study of any light cage's guidance and sensing characteristics in the THz regime is still missing. Furthermore, no studies so far have addressed the impact that any bends may have on photonic light cages, which is particularly important in the context of reconfigurability~\cite{cao2020additive}.

\begin{figure}[t!]
\centering
\includegraphics[width=\textwidth]{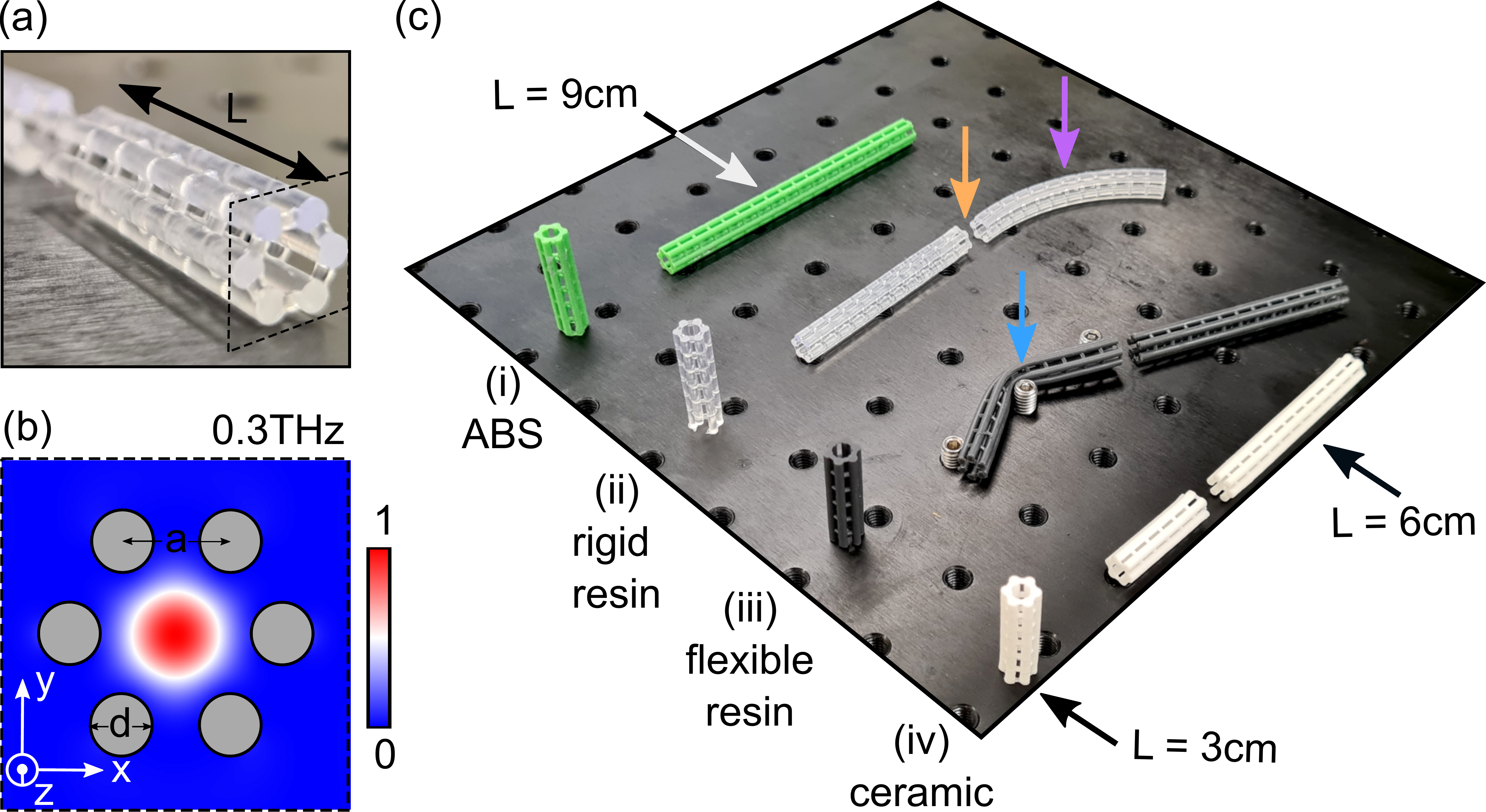}
\caption{3D printed Terahertz Light Cage (THzLC) device overview. We measure and characterize the terahertz transmission- and bend- losses of several 3D-printed THzLC modules. (a) Image of an example light cage, of length $L=3\,{\rm cm}$. (b) Example calculated normalized fundamental mode intensity at 0.3\,THz (window size: 1\,cm $\times$ 1\,cm), highlighting waveguide feature sizes and confinement characteristics. (c) Overview photograph of the THzLC modules considered here, composed of (i) ABS (ii) clear resin, (iii) flexible resin, and (iv) ceramic. We also show a bent clear resin light cage (purple arrow), which can be connected to other modules (orange arrow). In contrast, the flexible THzLC can be bent around arbitrary objects (blue arrow).}
\label{fig:concept}
\end{figure}

Here, we fabricate and experimentally characterize in detail the propagation losses, guided modes, and bend-loss characteristics of terahertz light cages (THzLCs), and demonstrate their versatility as waveguides and as sensors made out of a variety of materials, including examples capable of sensing at high temperatures. The straight- and bent- light cages are 3D printed up to lengths of 9\,cm, and bend radii down to 4\,cm, naturally lending themselves as modular elements of arbitrarily arranged assemblies. A picture of an example THzLC of length $L=3\,{\rm cm}$ is shown in Fig.~\ref{fig:concept}(a). As a first proof-of-concept study, our design is a scaled-up version of a previous design~\cite{jain2018hollow}: it consists of six, hexagonally arranged, high-aspect ratio strands (strand diameter: $d=2\,{\rm mm}$; center-to-center pitch: $a=3\,{\rm mm}$) connected by a thin bridge, as shown in Fig.~\ref{fig:concept}(a). By omitting the central strand of the hexagonal lattice, a hollow core is formed, which supports an antiresonant guided mode, shown in Fig.~\ref{fig:concept}(b). As noted, the THzLC allows direct access
to the hollow core via the open sections between transversally separated strands. The large fractions of modal fields within the hollow section (typical value: $\>99.9\%$), leading to diffractionless propagation, potentially over hundreds of wavelengths, while being in direct contact with the external environment~\cite{jain2018hollow}. 

To show the versatility of this kind of device, we study the terahertz transmission- and bend- characteristics of several nominally equivalent structures (in terms of cross-sectional geometric parameters), while varying the length, bend radii, and material. Each THzLCs is made from one of four different materials, using three distinct 3D printing procedures. Figure~\ref{fig:concept}(c) shows an overview of the different structures considered and their salient characteristics. Rigid light cages are constructed using three different polymers: (i) acrylonitrile butadiene styrene (ABS) (Filaform ABS, green); (ii) a clear resin (Formlabs Clear V4) (both (i) and (ii) are rigid); (iii) a flexible resin (Formlabs Flexible V2), which can be reversibly deformed; (iv) a rigid and robust ceramic (Baghdadite)~\cite{mirkhalaf2021personalized}. 3D-printed Baghdadite has recently attracted applications in the context of bone tissue engineering~\cite{lu2020baghdadite,no2021development}, and most importantly can withstand temperatures of at least 1400$^{\circ}{\rm C}$~\cite{mirkhalaf2021personalized}. Each of these materials would usually be considered to have prohibitively large losses for any THz applications (absorption: 10-100 dB/cm), but each can lend itself to form the cladding of anti-resonant hollow-core waveguides (propagation loss: 1-5\,dB/cm), because loss in the latter is dominated by radiation leakage due to antiresonant guidance, as opposed to material loss, as a result of the vanishingly small  overlap of antiresonant core modes with the solid strands.  

The rest of the paper is structured as follows. We first present the fabrication and terahertz loss measurements of the  THzLC and of the materials composing them; we then measure the terahertz near-field of the supported modes in two example rigid light cages, i.e., in a straight- and bent- configurations; next, we provide a detailed measurements and simulations of the bend losses of these light cages for bend radii below 10\,cm; finally, we present experiments presenting their versatility, and use them as either flexible components of modular assemblies, in-core sensors, or waveguide sensors operating at high temperatures, before concluding with some final remarks.

\section{Fabrication}

All light cages are printed from the same design template, produced by using Solidworks (MA, US).The ABS light cages, shown in Fig.~\ref{fig:concept}(c)(ii) were fabricated using a Funmat HT filament printer (Filaform ABS 1.75\,mm; print temperature: 220$^{\circ}{\rm C}$). The resin light cages are fabricated using a Formlabs 3B printer, using a clear rigid resin (Formlabs Clear V4, Fig.~\ref{fig:concept}(c)(ii)) and a flexible resin (Formlabs Flexible V2, Fig.~\ref{fig:concept}(c)(iii)). After printing, each resin light cage is first washed in IPA and cured for 15 minutes at 60$^{\circ}{\rm C}$. The ceramic light cage (Fig.~\ref{fig:concept}(c)(iv) is fabricated following the procedure described in Ref.~\cite{mirkhalaf2021personalized}. Briefly, a photosensitive ceramic resin was formulated by mixing 65wt\% of the ceramic (Baghdadite) powder with a 1:1 ratio of a photopolymer (Formlabs clear resin V4, NY, US) and a dispersant (Tween 20, Sigma Aldrich). A basic desktop printer (Formlabs 2, NY, US) was then used to print the geometries. Separately, we use mm-thick samples of the constituent materials to characterize their complex refractive index following the procedure outlined in Ref.~\cite{jepsen2019phase}

\section{Transmission experiments}

\begin{figure}[t!]
\centering
\includegraphics[width=1.0\textwidth]{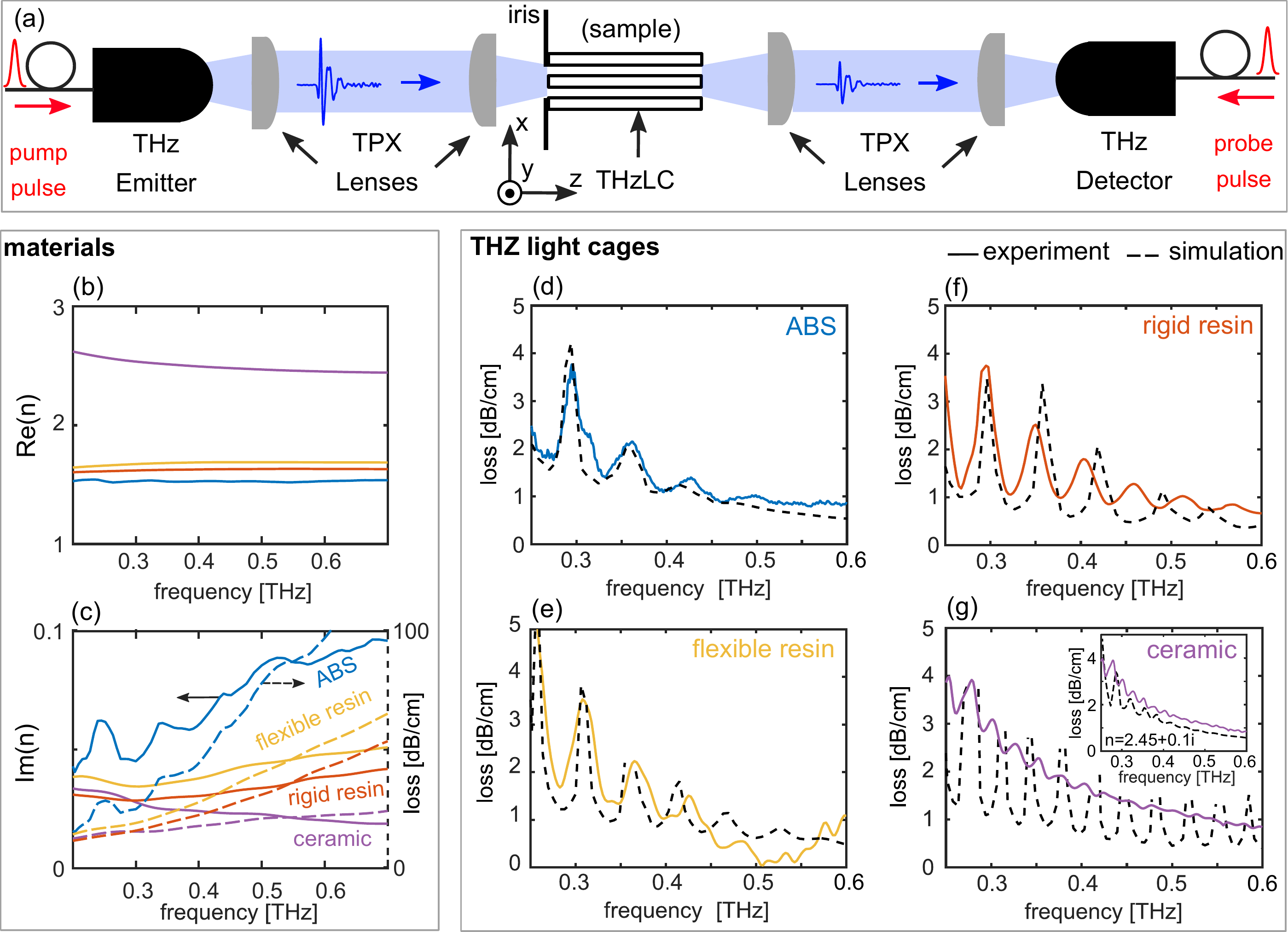}
\caption{Schematic of the fiber-coupled THz TDS experimental setup (Menlo TERAK15), used to measure waveguide loss and material refractive index. See main text for details. (b) Real and (c) imaginary parts of the refractive index of the materials used as the THzLC strands (left axes, solid lines). Blue: ABS; orange: rigid resin; yellow: flexible resin; purple: ceramic. Dashed lines (right axis) in (c): corresponding material loss (in dB/cm). Also shown is the associated measured propagation losses of the THzLCs composed of (d) ABS, (e) flexible resin, (f) rigid resin, and (g) ceramic. Dashed lines show the calculated loss profile for each geometry using COMSOL. Inset in (g) compares the measured loss with simulations, assuming a higher material loss via the refractive index as labelled.}
\label{fig:loss}
\end{figure}

Figure~\ref{fig:loss}(a) shows a schematic of the experimental setup used to measure the waveguide propagation loss and refractive index of each material. We use a commercially available THz-TDS System (Menlo TERAK15), which relies on THz emission from biased photoconductive antennas that are pumped by fiber-coupled near-infrared pulses (pulse width: 90\,fs; wavelength: 1560\,nm).  Polymethylpentene (TPX) lenses (Thorlabs TPX50) collimate and focus the beam towards the sample core, previously aligned using an iris. The THz field emerging from the THzLC is sampled as a function of the time delay of a fiber-coupled probe pulse on another photoconductive antenna, which forms the THz detector. Here we consider the the electric field polarized in $x$, using the sample orientation and reference frame shown in Fig.~\ref{fig:concept}(b); all waveguide bends are in the $xz$ plane. These choices are guided by preliminary calculations and prior experiments~\cite{bao2015dielectric,Stefani:21} indicating that such configurations have the lowest loss; a complete study of how the overall performance is affected by alternate choices of the polarization- and bend- directions is beyond the scope of this first proof-of-concept study.

Figures~\ref{fig:loss}(b) and \ref{fig:loss}(c) respectively show the measured refractive index and loss of each material used, obtained by measuring their respective transmitted THz pulses, and using the method outlined in Ref.~\cite{jepsen2019phase}. Subsequently, the THzLC attenuation was experimentally obtained by measuring the transmission through three straight waveguides (3--9\,cm lengths) with the same cross-sectional profiles, and fitting it for each frequency with an exponential function. The solid lines in Fig.~\ref{fig:loss}(d)--(g) show the measured loss of the THzLCs composed of different materials as labelled, as a function of frequency, showing regions of alternating low and high absorption corresponding to the strands' high (low) reflectivity at antiresonance (resonance)~\cite{jain2018hollow}. The behaviour and underlying physics is analogous to guidance using a single dielectric tube with wavelength-scale thickness~\cite{bao2015dielectric}; here the resonant reflections, rather than being caused  Fabry-Perot resonances of a thin dielectric tube, are due to the coupled resonances of the surrounding cylindrical strands~\cite{jain2018hollow} -- with the important additional feature that here the core region and the surrounding environment are directly connected. A comparison with calculated loss of the fundamental is plotted as a dashed line, showing good agreement. Our calculations use the Mode Analysis Package of the Finite Element Software COMSOL~5.3~\cite{comsol2017comsol}, which computes the propagation constants and mode profiles at each frequency, using perfectly matched layers (cylindrical coordinates) at the external-most boundary. All calculations use the experimentally measured complex refractive index shown in Fig.~\ref{fig:loss}(b),(c) for each respective THzLC strand, with $d=2\,{\rm mm}$ and $a=3\,{\rm mm}$ as per Fig.~\ref{fig:concept}(b). As a first experimental study of THz light cages, and to simplify the discussion of the underlying physics, our calculations consider only the fundamental mode. Note in particular that Fig.~\ref{fig:loss}(g) shows minor discrepancies between measured an calculated loss values, and an absence of any antiresonant oscillating features. We believe that this is a result of the fact that the ceramic 3D printing procedure introduces additional material losses: the inset of Fig.~\ref{fig:loss}(g) shows the same calculations assuming an imaginary part of the ceramic refractive index of 0.1, which shows better agreement with the simulations. This additional loss does not significantly hinder THzLC performance over the length scales considered here, adding only 1\,dB/cm of loss; reducing additional losses introduced by the printing process for ceramic materials will be the subject of future work.

Waveguide losses are low compared to material losses, but material losses arguably present some benefits for light cages: the resonances of the strands have low quality factor, with relatively sharp resonant/antiresonant transmission features present with lower loss materials (Fig.\ref{fig:loss}(e,f)) washed out with higher loss materials, resulting in smoother transmission characteristics (Fig.~\ref{fig:loss}(g)).

\begin{figure}[b!]
\centering
\includegraphics[width=1.0\textwidth]{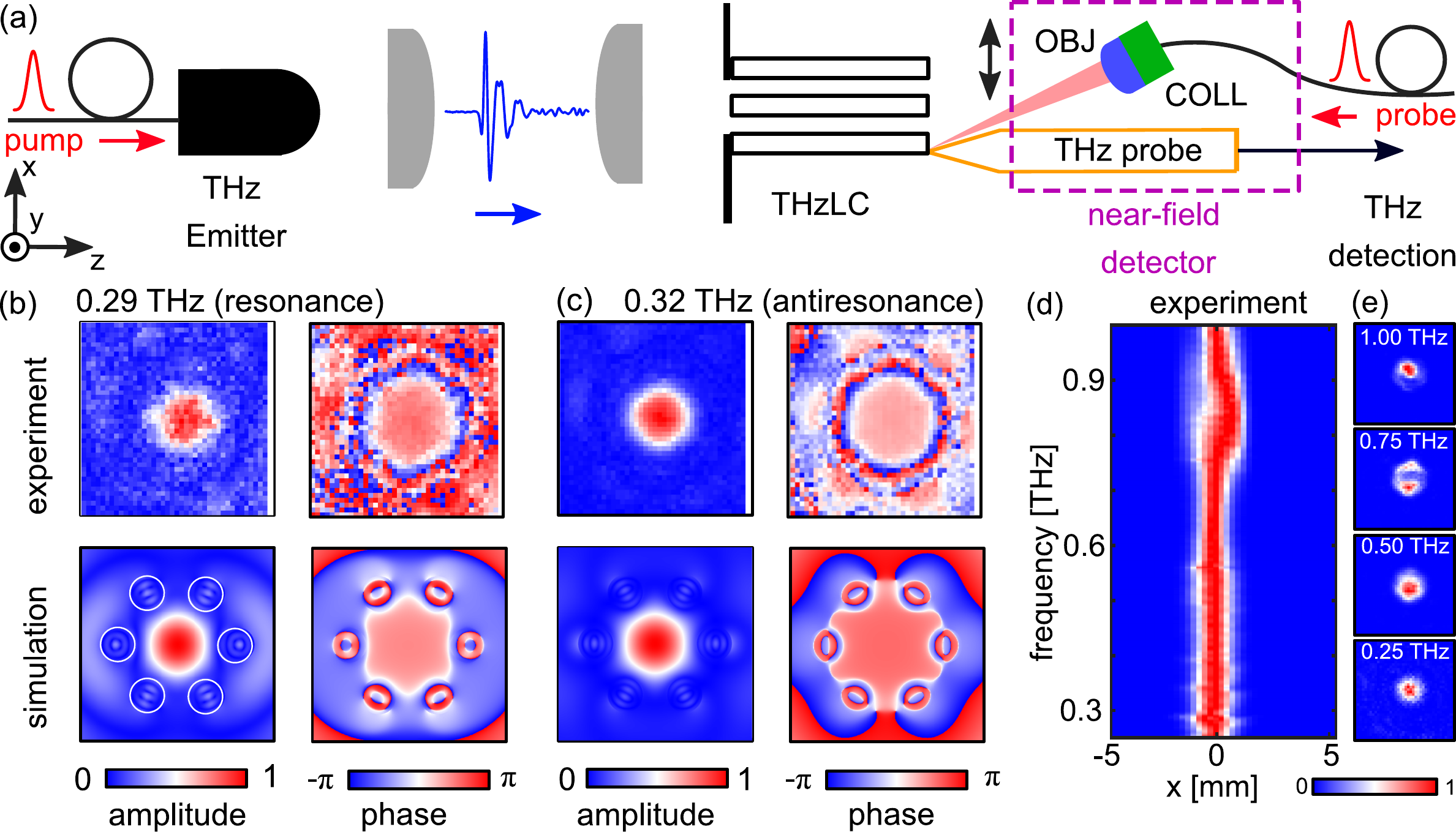}
\caption{Schematic of the modified fiber-coupled THz near-field TDS experimental setup (Menlo TERAK15), used to measure the electric field at the output of the waveguide. A moveable, fiber-coupled near-field detector module enables the measurement of the electric field at the output of a 6\,cm length rigid resin light cage. See main text for details. (b) Near-field measurements (top row) and simulations (bottom row) of the normalized amplitude (left) and phase (right) at the waveguide output, shown at a frequency of 0.29\,THz, corresponding to resonant guidance region (waveguide local loss maximum). (c) Same as (b), but at the antiresonant guidance region 0.32\,THz (waveguide local loss minimum). Window sizes: 1\,cm $\times$ 1\,cm. (d) Experimental frequency-dependent beam intensity profile along $x$ averaged in $y$ over the hollow core region, normalized at each frequency. (e) Example intensity mode images at 0.25\,THz spacing. Note that multiple modes are excited at higher frequencies. See Supporting Information for animations of (b),(c).
}
\label{fig:modes_straight}
\end{figure}

\section{Near field experiments}

We next measure the near-field emerging from the waveguides. 
Figure~\ref{fig:modes_straight}(a) shows a schematic of the near-field measurement system: it is the same as that of Fig.~\ref{fig:loss}(a), with the notable difference that the fiber carrying the probe pulse, which was previously directed to a detector in the far field, is now focused to a near-field THz probe  (Protemics TeraSpike TD-1550-X-HR), using a fiber collimator and objective. The entire unit (dashed line in Fig.~\ref{fig:modes_straight}(a)) forms a monolithic, moveable near-field detector module which raster scans the waveguide output in $x$ and $y$. The scans were performed on a $10 \times 10\,{\rm mm}^2$ area with lateral steps of 0.25\,mm, at a distance of about $300\,\mu{\rm m}$ from the end-face of the THzLC. Both emitter and detector are oriented to generate and detect linearly polarized fields in $x$, respectively. The electric near-field at output was thus directly measured as a function of time, at every $x$ and $y$ coordinate. One important feature of THz-TDS is that it directly maps the physical electric field, including both amplitude and phase information, so that the measured amplitude of the electric field at every frequency can then be immediately obtained from a Fourier transform~\cite{jepsen2011terahertz}. 

\subsection{Straight waveguides}

To reveal the guidance properties of the THzLC directly, we first consider a straight THzLC composed of rigid resin (length: 6\,cm), corresponding to Fig.~\ref{fig:concept}(c)(ii). The top row Figure~\ref{fig:modes_straight}(b) shows the  near field amplitude (left) and phase (right) measured at a representative frequency of 0.29\,THz, which corresponds to the region of (local) high waveguide loss -- see also Fig.~\ref{fig:loss}(f). The near field measurements reveal that, although most of the power is in the hollow core, some power can be found in the surrounding regions, due to resonant coupling to higher-order strand modes~\cite{jain2018hollow}. In contrast, in Figure~\ref{fig:modes_straight}(c), which shows the same measurements at 0.32\,THz,  most of the power is in the central core region, as a result of antiresonance guidance effects. Nevertheless, the measured phase in this window, shown in the top right of Fig.~\ref{fig:modes_straight}(c), distinctly reveals the existence of six light cage strands - showcasing the antiresonant guidance mechanism. These phase measurements in low-intensity regions are uniquely enabled by the high signal-to-noise ratio at these frequencies ($\sim$40\,dB). A comparison with the simulation of the intensity- and phase- of the fundamental mode profile, shown in the bottom row Fig.~\ref{fig:modes_straight}(b) and \ref{fig:modes_straight}(c), shows good agreement of the salient experimental features.

Figure~\ref{fig:modes_straight}(d) shows the average intensity distribution in the central core region as a function of $x$ and frequency, confirming that the majority of the power is contained in the hollow core region in the entire sub-THz band. Although the waveguide is multi-moded, our measurements show coupling into a single mode, with diffractionless guidance of the fundamental mode up to about 0.5\,THz. Figure~\ref{fig:modes_straight}(e), shows the modal intensity images at 0.25\,THz frequency spacing: the power distribution at 0.75\,THz and 1.00\,THz reveal that multiple modes are excited, but at 500\,GHz and below the fundamental mode dominates. 

\subsection{Bent waveguides}

We now investigate the properties of bent light cages under typical experimental conditions, with an eye on developing convenient and reconfigurable THzLC circuits to guide THz fields at will -- for example, either by assembling rigid elements~\cite{cao2020additive}, or reconfigurably bending flexible elements~\cite{Stefani:21}. We begin by considering a rigid resin light cage, of length $L=6\,{\rm, cm}$ and bend radius $R_b = 5\,{\rm cm}$, shown by the purple arrow in Fig.~\ref{fig:concept}(c). Analogously to previous approaches~\cite{setti2013flexible, bao2015dielectric, Stefani:21}, the bend radius $R_b$ is the ratio between the waveguide length and the angle (in radians) swept by the {\em external} profile of each waveguide. Due to the cm-scale waveguide diameter and bend radii, such resulting bend radii are approximate, and can be considered as upper limit estimates. 

\begin{figure}[t!]
\centering
\includegraphics[width=1.0\textwidth]{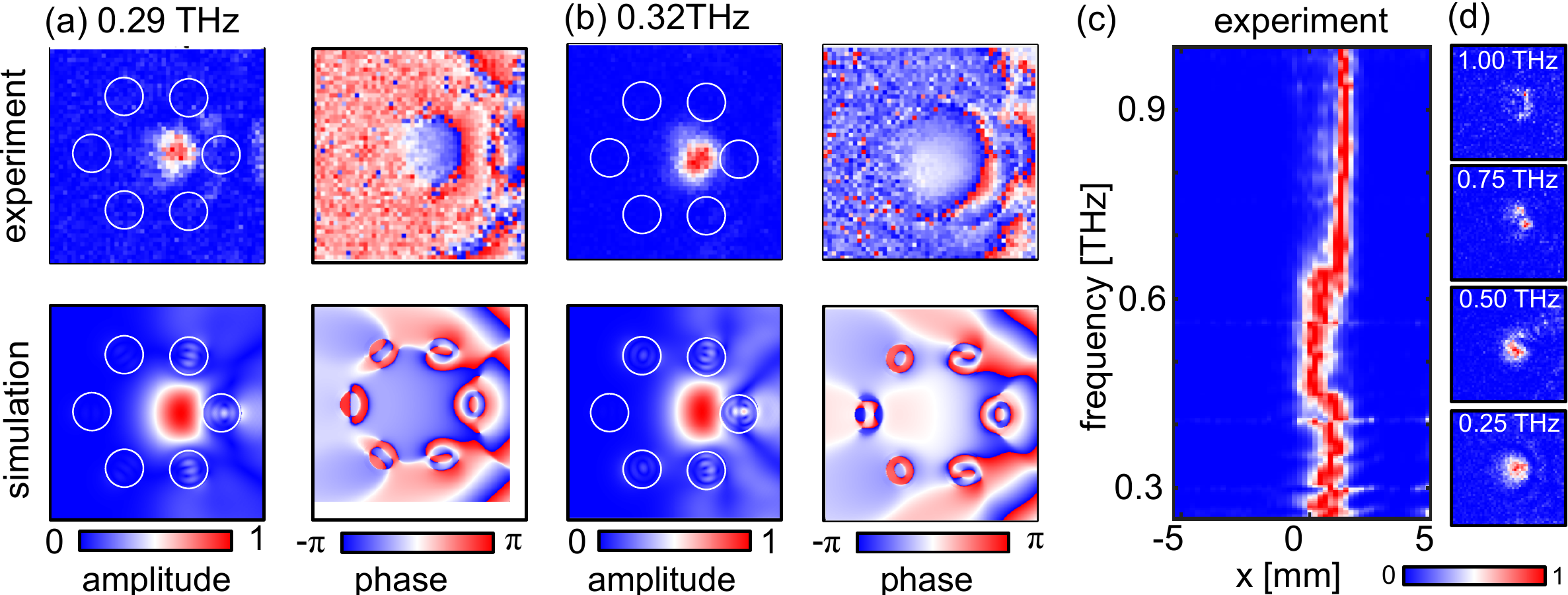}
\caption{(a) Near-field measurements (top) and simulations (bottom) of magnitude and phase of a curved THzLC (rigid resin; bend radius: 5\,cm) at 0.29\,THz and (b) at 0.32\,THz, for comparison with Fig.~\ref{fig:modes_straight}. Window sizes: 1\,cm $\times$ 1\,cm. (c)  Experimental frequency-dependent beam intensity profile along $x$ averaged in $y$ over the hollow core region, normalized at each frequency. (d) Example intensity mode images at 0.25\,THz spacing. See Supporting Information for animations of (a),(b). 
}
\label{fig:modes_bent}
\end{figure}

The top row of Figure~\ref{fig:modes_bent}(a) and ~\ref{fig:modes_bent}(b) show the  near field amplitude (left) and phase (right), at frequencies of 0.29\,THz and 0.32\,THz respectively for ease of comparison with Fig.~\ref{fig:modes_straight}. One may clearly observe a strong deformation of the measured field, in the direction of the waveguide bend, as is typical for any strongly bent waveguides~\cite{yuan2005characterization}. Similar measurements have been performed in the far field for several terahertz waveguides~\cite{doradla2012characterization, navarro2013terahertz, bao2015dielectric, Stefani:21}, and in the near field for planar waveguides in the longitudinal plane~\cite{yuan2005characterization, bozhevolnyi2006near}; to our knowledge however, this is the first transverse near-field measurement of a the complex field cross-section emerging from a waveguide with extreme bends. 
 
To interpret these results, we compare our experiments with numerical models. We adapt out finite element mode solver to model an equivalent refractive index profile $n_b$ for the bent THzLCs, given by~\cite{heiblum1975analysis}

\begin{equation}
n_b(x,y) = n(x,y) \exp(-x/R_b),
\label{eq:nb}
\end{equation}
where $x$ is the (horizontal) bending direction, and $n(x,y)$ is the cross-sectional refractive index distribution of the straight waveguide considered. The resulting amplitude- and phase- profiles at these frequencies is shown in the bottom row of Fig.~\ref{fig:modes_bent}(a) and \ref{fig:modes_bent}(b). The simulations and experiments both show that these bends induce a radiative leakage of the field outside the waveguide, which here is the dominant loss mechanism. 

\begin{figure}[t!]
\centering
\includegraphics[width=1.0\textwidth]{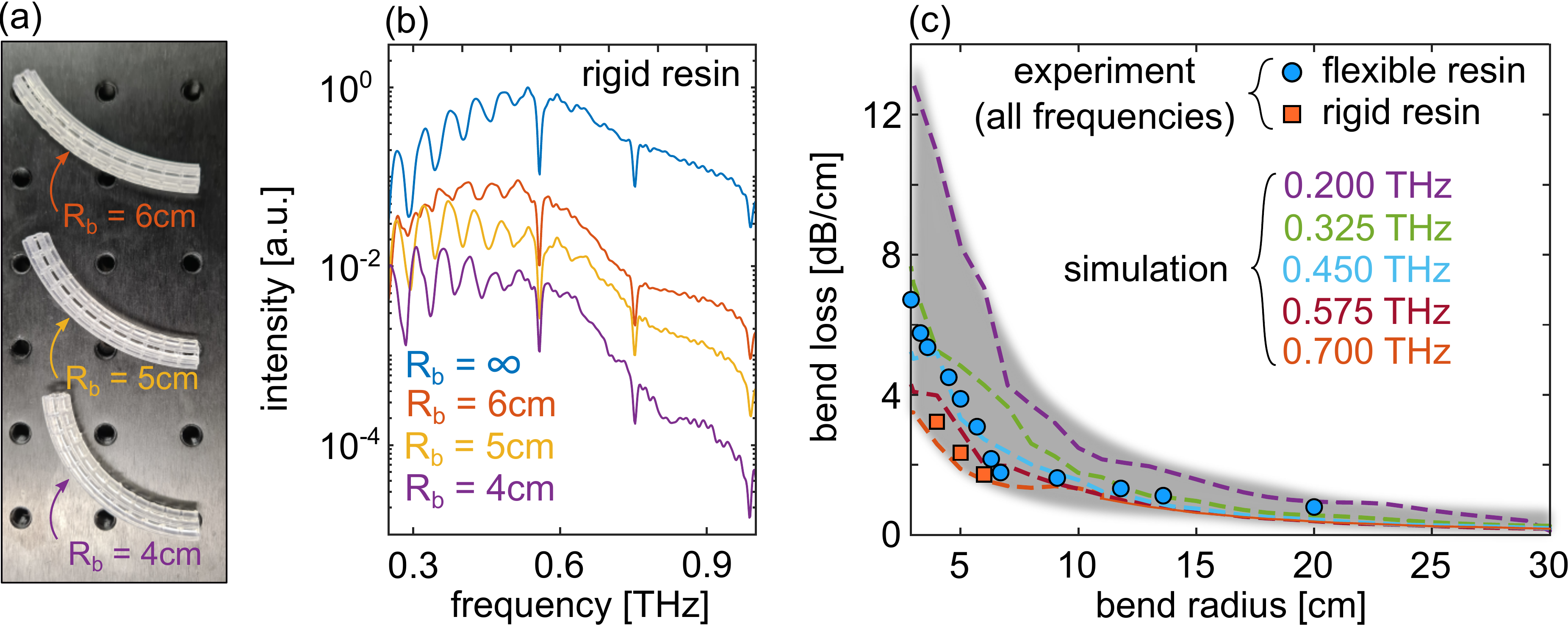}
\caption{(a) Photograph of the three curved rigid light cages of decreasing bend radius (ORANGE: 6\,cm; yellow: 5\,cm; purple: 4\,cm). (b) Associate transmitted intensity for each THzLC, using the same colour coding as (a). The blue line shows the transmission through a straight light cage. (c) Total bend loss in the range (0.2-1\,THz) for the rigid light cage (orange squares) and flexible light cage (blue circles) as a function of bend radius. Dashed lines show the calculated bend loss (COMSOL). Grey shaded area is a guide to the eye encompassing the bend loss range.}
\label{fig:bend_loss}
\end{figure}

To investigate this further, we first consider three curved rigid resin light cages, shown in Fig.~\ref{fig:bend_loss}(a), of decrasing bend radius (here: $R_b = 6\,{\rm cm}$ (orange), 5\,cm (yellow) and 4\,cm (purple)), and repeat the transmission experiments for each one. Note that, in the context of cm-scale terahertz waveguides, the bends presented here as typical are in fact quite strong: bending a 1\,cm diameter waveguide to a 4\,cm bend, for example, is analogous to bending a standard 125\,$\mu$m diameter fiber to a 0.5\,mm bend radius. The fiber-coupled THz system allows us to align the hollow core at input- and output- at the same nominal location relative to the emitter and detector, respectively. The resulting raw transmission spectra in each case are shown in Fig.~\ref{fig:bend_loss}(b), including the spectrum for the straight waveguide. We note that, as an overall trend, the transmitted power drops by 10-30\,dB as the bend radius decreases, with higher frequencies having larger bend losses. The oscillations at lower frequencies, associated with regions of low- and high- leakage loss as per Fig.~\ref{fig:loss}(f), do not have the local minima and maxima at the same frequency. A similar effect was previously observed for enclosed polymer tubes~\cite{bao2015dielectric}, and can be explained by inspecting the refractive index profile of the conformal map (Eq.~\eqref{eq:nb}): upon bending, effective refractive index changes in the strands shift the resonances/antiresonances. Furthermore, significant changes in the strand diameter could be occurring along the device length, due to the relatively small bend radii (relative to the wavelength.)

As a result, it is challenging to unambiguously quantify the bend loss at each specific frequency. Instead, we experimentally quantify the overall power transmitted by the terahertz pulse, and compute the degree to which bending the waveguide introduces additional losses to the overall transmission per unit length by defining the bend loss per unit length as

\begin{equation}
\alpha_{\rm bend}^{\rm exp} {\rm [dB/cm]} = -10 \log_{10} \left(I_{\rm bent}/I_{\rm straight}\right)/L,  \label{eq:bl}
\end{equation}
where $I_{\rm straight}$  and $I_{\rm bent}$  respectively represent the {\em total} transmitted intensity integrated across frequencies for a straight- and bent- waveguide of length $L$ in cm.  The orange squares in Fig.~\ref{fig:bend_loss}(c) plots $\alpha_{\rm bend}^{\rm exp}$, taking the total (integrated) intensity between 0.2-1\,THz. We find that the total bend losses are in the range of 2--4\,dB/cm, which is comparable to previous results in rigid and flexible enclosed tubes~\cite{bao2015dielectric, setti2013flexible, Stefani:21}, despite there  being  no continuous boundary between the mode and the outside air in this geometry. 

To verify the validity of our experiments, we perform finite element modal simulations in COMSOL using the refractive index mapping of Eq.~\ref{eq:nb}, for a few representative frequencies, and for a wide range of bend radii ($R_b = 3-100\,{\rm cm}$). For each frequency, the complex effective index of the calculated fundamental mode at each value of $R_b$ yields the simulated bend loss per unit length via

\begin{equation}
\alpha_{\rm bend}^{\rm sim} {\rm [dB/cm]} = -10 \log_{10} \left[ \exp \left( 2 k_0 \Im{\rm m}(\Delta n_{\rm eff} )  \times 1\,{\rm cm}\right)\right],  \label{eq:bl_sim}
\end{equation}
where $k_0 = 2\pi/\lambda$ is the free space wave vector in units of $\rm cm^{-1}$, $\Delta n_{\rm eff} = n_{\rm eff}^{\infty} - n_{\rm eff}^{R_b}$, and $n_{\rm eff}^{R_b}$ is the effective index calculated for  a bend radius $R_b$. The resulting calculated bend losses of the fundamental mode of the THzLC at frequencies between 0.2--0.7\,THz are shown as dashed lines Fig.~\ref{fig:bend_loss}(c). The grey shaded region encompasses the range of calculated losses between 0.2-0.7\,THz -- where most of the power is contained. The simulations overlap well with our experiments, and confirm that our experimental results for the bend loss are within the expected range. Note however that the calculated bend loss of the  fundamental mode at higher frequencies (e.g., 0.7\,THz) is smaller than that at lower frequencies (e.g., at 0.2\,THz), a trend that is opposite to that of our experiments. This is most likely because, in practice, higher modes are excited at high frequencies (and bending itself can lead to mode coupling), increasing the overall bend loss, and making direct comparisons with single-mode simulations challenging. 

The flexible, reconfigurable resin waveguides, shown in Fig.~\ref{fig:concept}(c)(iii), allow us to repeat these experiments using a wider range of $R_b$ while using the same sample. Custom-built THzLC sample holders, combined with the fiber-coupled THz system, ensure that the input- and output- conditions are left unchanged when these flexible THzLCs are bent~\cite{Stefani:21}. The blue circles in Figure~\ref{fig:bend_loss}(c) show the experimental results: the bend losses appear to be overall slightly higher, while still within the range suggested by our numerical simulations. 

\section{Applications}

Having characterized both straight- and bent- THzLC modules individually, we now turn our attention to potential applications, presenting three key separate but complementary experiments to illustrate the opportunities enabled by these devices: (1) modular, flexible, reconfigurable waveguide assemblies, which thus allow to re-route diffractionless free-space beams without additional free-space optics; (2) in-core- terahertz sensing, which is unique to the light cage geometry; and (3) high-temperature terahertz guidance and sensing, enabled by the use of ceramics as the device material. 

\begin{figure}[t!]
\centering
\includegraphics[width=0.8\textwidth]{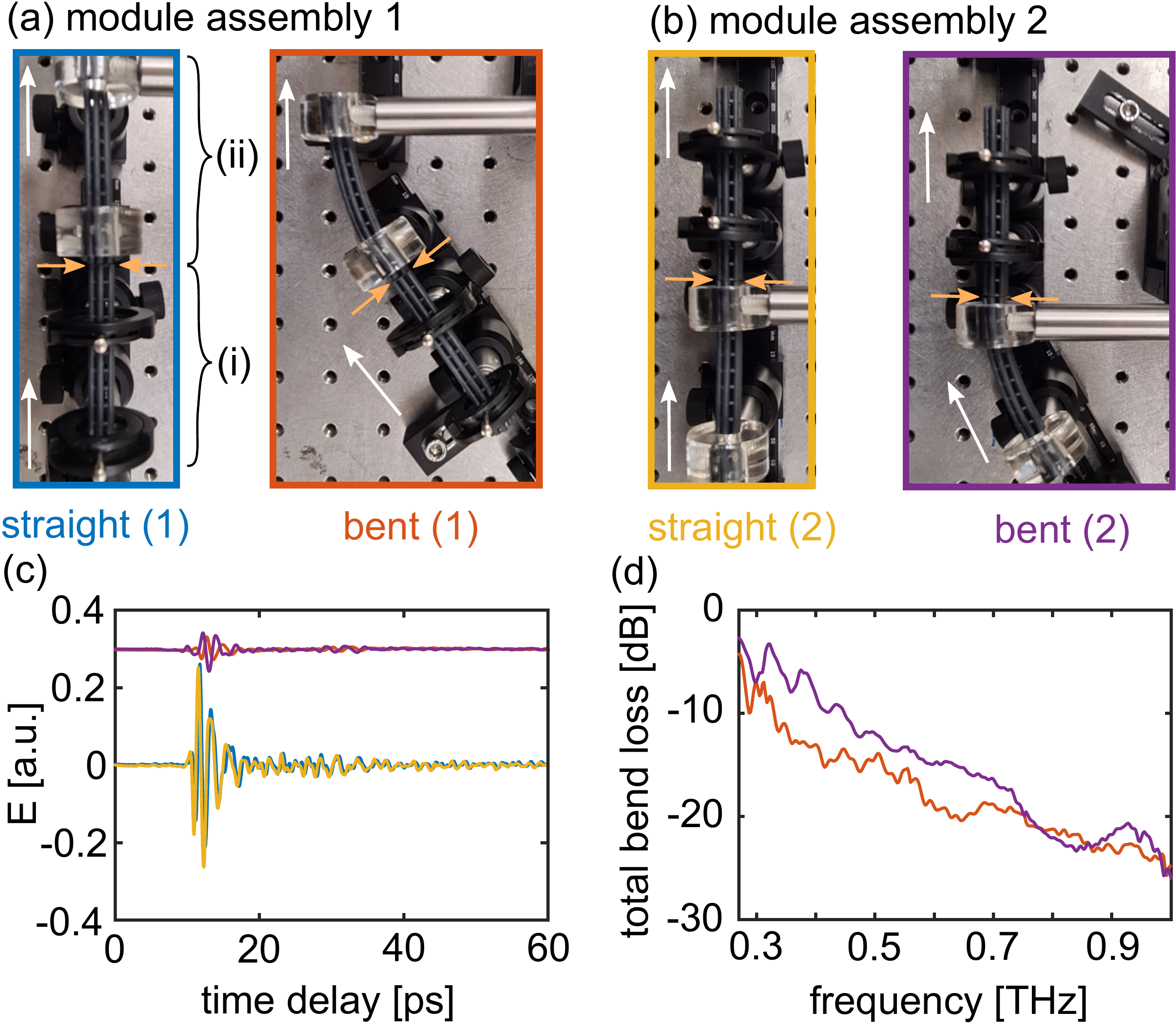}
\caption{Summary of experiments demonstrating THzLC modularity, flexibility, and reconfigurability. (a) Transmission measurements are performed on two straight THzLCs (i) and (ii) placed in series (blue frame); the experiment is repeated after bending THzLC (ii) (orange frame). (b) The same experiment is repeated in a striaght configuration (yellow frame), and in a configuration where THzLC (i) is bent by nominally the same angle (purple frame). The coloured arrows highlight the location where the two separate modules join. White arrows indicate the direction of propagation. (c) Associated transmitted electric field in all configurations, using the same colour code as (a). The orange and purple curves are vertically offset by +0.3 for clarity. (d) Frequency-dependent bend loss in each configuration.}
\label{fig:modular}
\end{figure}

\subsection{(1) Modular flexible assemblies}

We begin by considering flexible resin THzLC modular assemblies composed of two 6\,cm-elements placed in series, where either the first- or second- of the two elements is arbitrarily bent. Two module assemblies, composed of THzLC (i) and (ii), are shown in Fig.~\ref{fig:modular}(a) (blue) and ~\ref{fig:modular}(b) (yellow). Their respective transmitted terahertz electric field pulse measurement is shown as blue- and yellow- lines in Fig.~\ref{fig:modular}(c), showing that they are comparable, confirming that the coupling conditions are similar in each configuration. Subsequently, each module assembly is mechanically bent by nominally the same amount, while maintaining the input- and output- alignment; the resulting experimental configurations are shown in Fig.~\ref{fig:modular}(a) (orange) and ~\ref{fig:modular}(b) (purple), and their respective transmitted THz pulses are shown as the orange- and purple- curves in Fig.~\ref{fig:modular}(c). Note the drop in amplitude, which we quantify at each frequency and for both bent configurations. Figure~\ref{fig:modular}(d) shows the total intensity lost in each bent configuration with respect to each straight configuration. As one would expect from our data of Fig.~\ref{fig:bend_loss}(c), the total drop in intensity is between 5--25\,dB over the entire assembly length, which is predominantly due to the 6\,cm bent region. We find that higher frequencies experience more loss, most likely due to the excitation of higher oreder modes in each waveguides in that region. Note that a number of factors can influence the overall loss, including a mode mismatch between straight- and bent- WGs (see Fig.~\ref{fig:modes_straight} and Fig.~\ref{fig:modes_bent}), and the waveguide profile resulting from the specific mechanical bend, which can influence small local changes in the local bend radius and thus the overall transmission, especially when $R_b<10\,{\rm cm}$ as is the case here.

Our experiments show that transmission losses depend on the bends and exact assembly of modules - but remain at a level that could readily  be normalized out for many applications. However, in spectrally sensitive applications it will be important to carry out the normalization {\em in situ}, in the final THzLC configuration.

\begin{figure}[t!]
\centering
\includegraphics[width=0.7\textwidth]{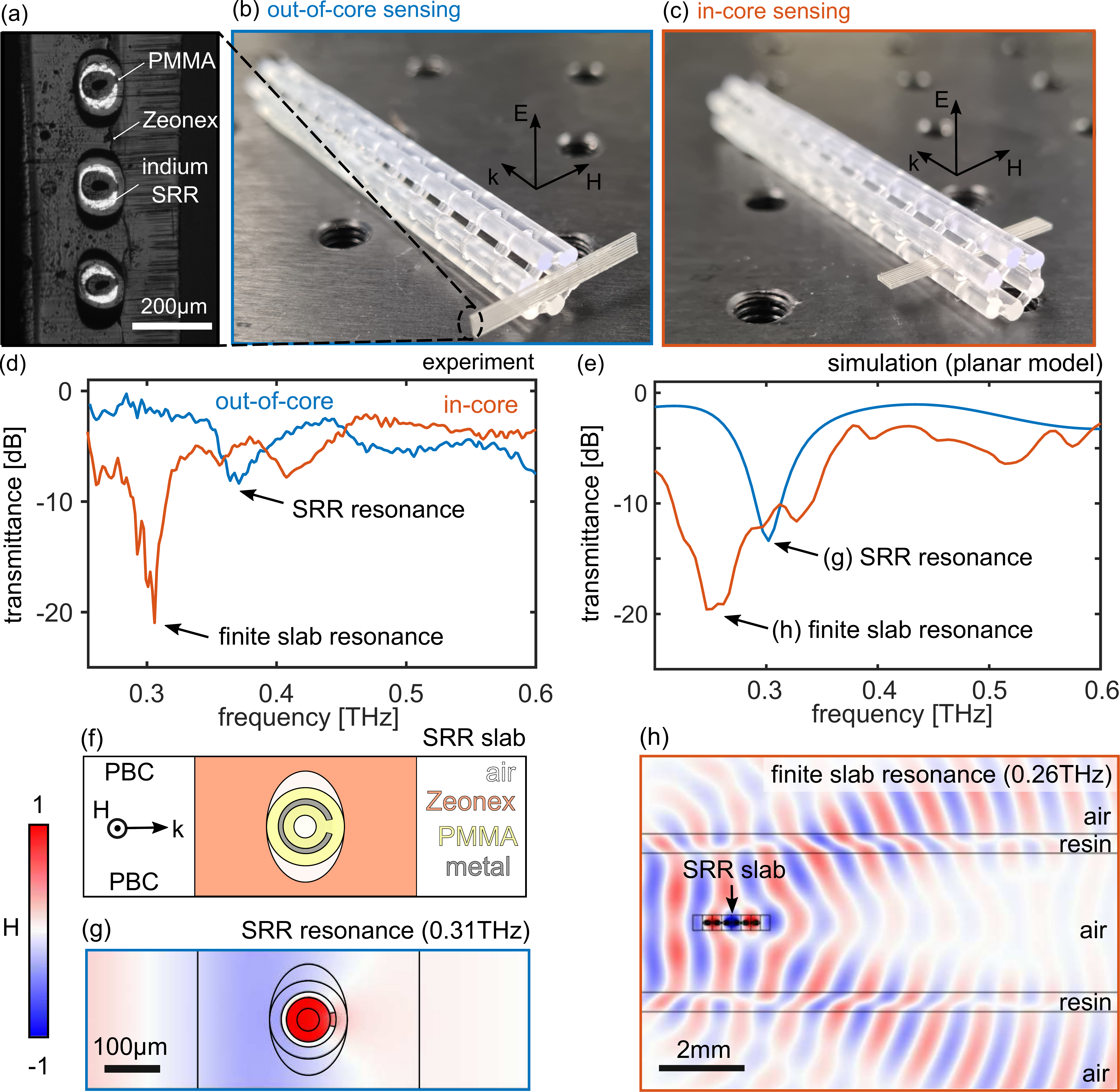}
\caption{Summary of experiments demonstrating out-of-core and in-core sensing capabilities. (a) Microscope image detail of the magnetic resonator array, composed of an indium slotted cylinder in a PMMA jacket, contained in a Zeonex slab~\cite{tuniz2011stacked}.  (b) Photograph of a THzLC with a slotted-cylinder resonator metamaterial slab placed out-of-core, and (c) in-core. In both cases the strongest magnetic field $H$ component of the excited core mode is parallel to the cylinder axis, achieved by appropriately rotating the emitter and detector antennas. (d) Transmittance in the out-of-core (blue) and in-core (orange) configuration. Note two distinct primary resonances in each case, which we attribute to the slotted-ring resonator (out-of-core) and the finite slab (in-core), as confirmed by (e) transmittance calculations using a planar mode. (f) Modelled geometry to identify the SRR resonance, using an idealized case of the central cylinder in (c). In this case, periodic boundary conditions (PBCs) are used to model the multi-cylinder slab in (c), as per Ref.~\cite{tuniz2011stacked}. (g) Calculated magnetic field distribution at the characteristic resonance of the sub-wavelength SRR~\cite{tuniz2011stacked}. (h) Calculated magnetic field distribution for a planar SRR slab slab placed inside an antiresonant slab waveguide. Note that the field scattered out of the core and into free space, strongly reducing the transmission at slab resonance. }
\label{fig:metamaterial}
\end{figure}

\subsection{(2) In-core sensing}

We now show that this structure allows to detect resonant features both outside \emph{and inside} the core of the waveguide, harnessing the unique characteristics of the THzLC. Most THz sensing applications rely on the observation and monitoring of characteristic molecular resonances~\cite{seo2020terahertz}; to perform a controlled experiment, here we use a slab containing resonant ``meta-molecules'', composed of slotted ring resonators (SRRs). Fig.~\ref{fig:metamaterial}(a) shows a detailed microscope image of the cross section of the SRRs array used. As previously reported~\cite{tuniz2011stacked}, it is formed by an array of six sub-wavelength slotted indium cylinders inside a PMMA jacket, which form an array inside a Zeonex slab (slab area: $0.4 \times 1.5\,{\rm mm}^2$; slab length: 30\,mm). A previous experiment~\cite{tuniz2011stacked} placed a single slab of this sample in the focus of a THz 1\,mm beam, such that most of the beam overlapped with the sample. It was found that this structure possess a magnetic resonance at 0.35\,THz when the magnetic field is aligned parallel to the cylinder axis. In the present context, an analogous experiment can be repeated by placing the slab at the output of a THzLC as illustrated in the photograph in Figure~\ref{fig:metamaterial}(b) (here termed ``out-of-core sensing''.) This kind of experiment can, in fact, be readily performed by placing the sample at the output of any cm-scale THz waveguide: for example, an extension of this technique in the context of bendable terahertz fibers was used to perform terahertz endoscopy~\cite{lu2014terahertz}. The THzLC, however, provides an additional avenue, as illustrated in Fig.~\ref{fig:metamaterial}(c): the ability to place the sample \emph{directly in the core} of the waveguide (here termed ``in-core sensing''). In this specific case, the sample is self-aligned with the center of the waveguide; more broadly, this technique guarantees that any changes in the transmission must originate from the core. Due to absence of continuous confining boundary, and the core's multi-modedness and  low numerical aperture, scattering sources produce radiation leakage directly into the surroundings, dropping the overall waveguide transmission, providing strong resonant signatures that can uniquely be attributed to elements in the core. 

To illustrate this point, we perform transmittance experiments using the out-of-core and in-core configurations. The antenna and detector are rotated so that the magnetic field is parallel to the slotted cylinders when the sample is arranged as in Fig.~\ref{fig:metamaterial}(b) and \ref{fig:metamaterial}(c), as required~\cite{tuniz2011stacked}. The blue and orange lines in Fig.~\ref{fig:metamaterial}(d) show the transmittance in each respective case. For the out-of-core case, we observe a $\sim 10\,{\rm dB}$ resonant dip at 0.35\,THz, corresponding to the previously reported magnetic resonance of the SRR~\cite{tuniz2011stacked}. For the in-core case, we measure a dominant resonant dip at lower frequency of 0.3\,THz, accompanied by two other less prominent resonances at 0.35\,THz and 0.4\,THz. In order to understand this result, we simulate this system using a 2D propagation model, which captures the fundamental physics with relatively fast computation times. A schematic resonator used in the model is shown in Fig.~\ref{fig:metamaterial}(f), which is an idealized version of the central resonator shown in Fig.~\ref{fig:metamaterial}(a). 

Analogously to earlier treatments~\cite{tuniz2011stacked}, the out-of-core transmission is modelled using plane wave port input with periodic boundary conditions. An example simulation of the magnetic field at resonance is shown in Fig.~\ref{fig:metamaterial}(g). The resulting transmittance as a function of frequency is shown as a blue line in Fig.~\ref{fig:metamaterial}(e), showing the characteristic resonant feature observed in the out-of-core experiment.  

To capture the physics of the in-core sensing capabilities, our numerical analysis model replicates the experimental condition of Fig.~\ref{fig:metamaterial}(c) by considering a 1D slab containing six SRRs inside an antiresonant waveguide, as shown in Figure~\ref{fig:metamaterial}(h). We consider an array of six resonators inside a one-dimensional waveguide, formed an 5\,mm air core surrounded by two 0.5\,mm slabs ($n=1.6+0.03i$), forming a simple 1D antiresonant slab waveguide. The transmittance as a function of frequency is shown as an orange line in Fig.~\ref{fig:metamaterial}(e). Despite the simplicity of this in-core-sensing model, we observe the same salient features as our experiment: (i) a strong resonance at a lower frequency than the out-of-core case -- which, via the associated field plots in Fig.~\ref{fig:metamaterial}(h), we can attribute to a Fabry Perot type resonance of the finite slab; and (ii) additional weaker resonance emerging at slightly higher frequencies, which are due to the SRR- and higher-order- slab resonances, respectively. Note that this kind of experiment, were it performed at the focus of a free-space beam using two lenses, would be challenging to perform and difficult to interpret. For example, at a frequency of 0.3\,THz ($\lambda=1\,{\rm mm})$, the fitted Gaussian $1/e^2$ beam waist radius of the intensity profile equivalent to one emerging from the THzLC is $w_0 = 1\,{\rm mm}$ (Fig.~\ref{fig:modes_straight}(d)), corresponding to a Rayleigh length of only $L_R = \pi w_0^2 / \lambda = 3.14\,{\rm mm}$, barely matching the sample size. As a result, performing this experiment in free space would  require precise alignment and identification of the exact location of the focus in three dimensions; furthermore, the lenses collecting the scattered light could themselves capture some of the scattered power at the focus depending on alignment, and features observed would be highly dependent on alignment and be open to interpretation. 
Here, by collecting the light at the output of a 60\,mm waveguide supporting diffractionless propagation, the scattering source can be anywhere inside the waveguide, which in this case is also self-aligned with the central part of the cage core. Thus, the physical location of the resonant scattering and light collection are separated, offering both convenience and ease of interpretation of any spectral features.

\begin{figure}[t!]
\centering
\includegraphics[width=\textwidth]{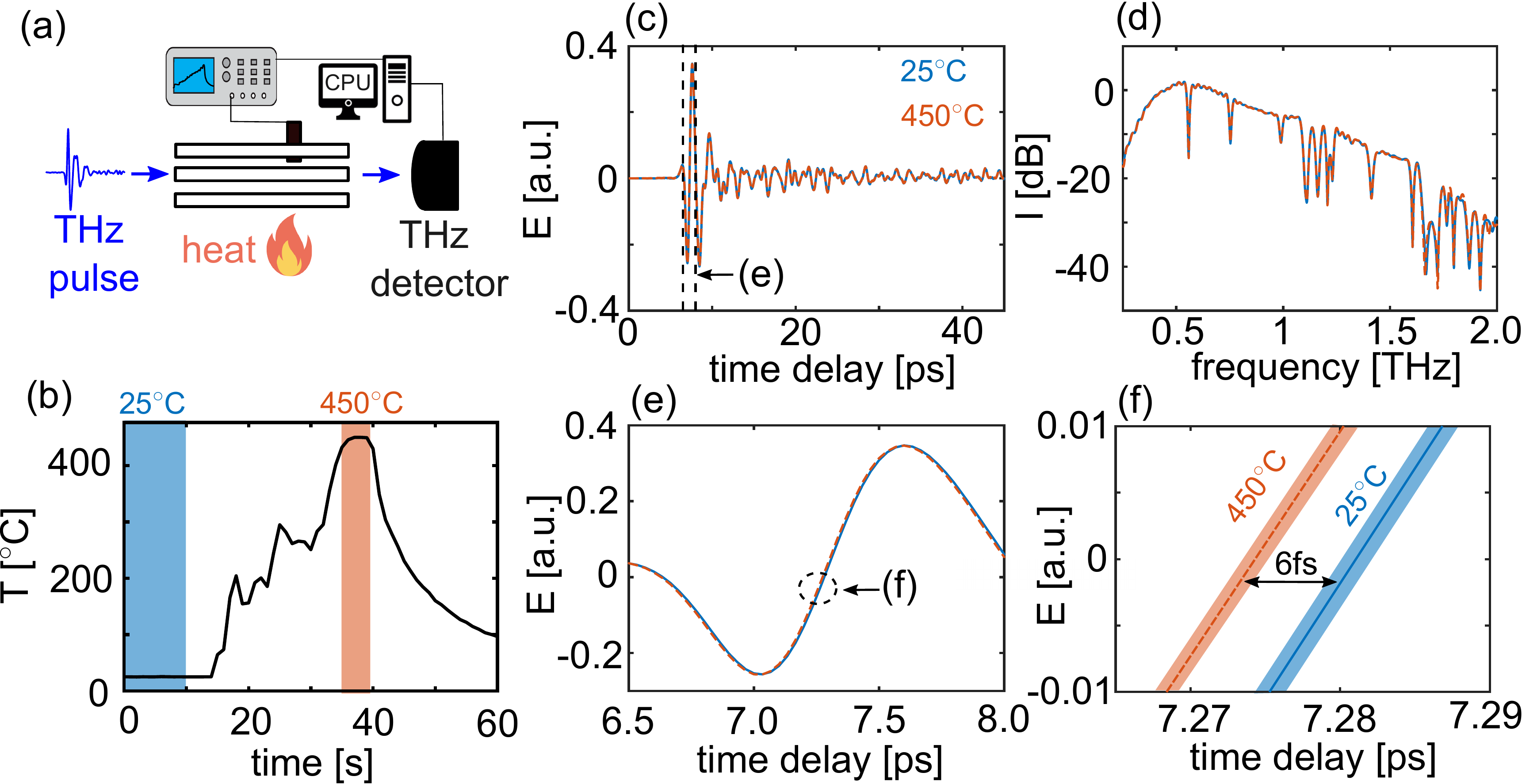}
\caption{Summary of high-temperature THzLC measurements. (a) Experimental schematic: the transmission experiment is modified to include a calibrated thermocouple in contact with the ceramic THzLC, and a heat source. A terahertz detector and oscilloscope respectively  measure the terahertz pulse and temperature every second. (b) Temperature profile as a function of time. Blue- and orange- shaded regions highlight the lowest- and highest- temperatures of 25\,$^\circ$C and 450\,$^\circ$C, respectively. Also shown are (c) the measured terahertz pulse and (d) the associated spectrum, plotted on a log scale, at 25\,$^\circ$C (blue line) and  450\,$^\circ$C (orange dashed line), showing comparable transmission spectra. (e) Detail of the terahertz pulse in (c), time-averaged over the low- and high- temperature regions highlighted in (b). (f) Detail of the terahertz pulse in (e), showing a characteristic 6\,fs time delay~\cite{mittleman1998gas}. Shaded regions encompass the standard deviation.}
\label{fig:temperature}
\end{figure}

\subsection{(3) High temperature guidance and sensing}

We now demonstrate the potential of using the ceramic light cages as sensors operating at higher temperatures than those allowed by conventional 3D printing polymer-based materials, which typically deform or melt at temperatures close to 150--200\,$^\circ$C. Operating waveguides at extremely high temperatures has applications in optical sensing within harsh environments~\cite{ghosh2019high}, which in the context of terahertz radiation allows, for example, remote monitoring of gases (e.g., H$_2$O, CO) inside gasifiers~\cite{bidgoli2014terahertz} and furnaces~\cite{song2015high}, with implications for industrial process control. Our modified experimental setup for measuring the temperature-dependent terahertz spectrum is shown in Fig.~\ref{fig:temperature}(a). The far-field THz-TDS light-cage transmission setup of Fig.~\ref{fig:loss}(a) is modified to include a thermocouple in contact with a ceramic light cage ($L=6\,{\rm cm}$), and connected to an oscilloscope via an amplifier. This setup thus enables us to measure the temperature of the THzLC while the terahertz pulse is passing through it (refresh rate: 1\,Hz). As a readily available heat source, we use the butane flame from a common household lighter, held directly underneath the light cage for $\sim$30\,s. Figure~\ref{fig:temperature}(b) shows the time-dependent temperature profile as the heat source is turned on: the starting temperature is 25\,$^\circ$C, and reaches a maximum of 450\,$^\circ$C. The associated terahertz pulse and spectrum at the lowest- and highest- temperatures measured are shown in Fig.~\ref{fig:temperature}(c) and \ref{fig:temperature}(d) respectively: they are similar, and confirm that the ceramic light cage's guidance is not significantly affected at such high temperatures. In addition, we detect a small phase shift of the terahertz pulse at high temperatures: Figs.~\ref{fig:temperature}(e) and \ref{fig:temperature}(f) respectively show the terahertz pulse, time-averaged over the shaded regions in Fig.~\ref{fig:temperature}(b), over smaller time windows. The shaded  regions in Fig.~\ref{fig:temperature}(f) encompass the standard deviation. We observe a 6\,fs in the time shift, in agreement with expectations~\cite{mittleman1998gas}, and which we attribute to a reduction of the air's refractive index in the cage core, which causes the pulse to reach the detector sooner. Note that the maximum temperature reached is well above the melting temperature of plastics and resins, but significantly below the 1400\,$^\circ$C which the ceramic light cages can potentially withstand, due to inherent difficulties associated with installing and safely using high-temperature sources within our university's terahertz laboratory. Nevertheless, we believe the present experiments are a valuable proof-of-concept starting point for showing the viability of using ceramics as the cladding material for modular antiresonant terahertz waveguides and sensors in harsh environments~\cite{ghosh2019high} -- which might include simple tubes, but which in the specific case of the THzLC also allows direct access to the guiding core as discussed.

\section{Conclusions}

Light cages can confine and guide terahertz radiation, while offering direct lateral access to the guided mode. With propagation losses in light cages largely independent of the materials used, we demonstrated that additive manufacturing is a versatile fabrication tool that is well suited for the required geometries, including for modular, flexible and  high-temperature resistant waveguides.  

Our near-field measurements revealed the characteristic leaky resonant- and antiresonant- modes involved in light guidance; most remarkably, the measurements in a bent configuration showed excellent agreement with that predicted by solving for the fundamental mode using a conformal map~\cite{heiblum1975analysis} -- an aspect which has, to the best of our knowledge, eluded measurement until now.  

Our experiments reveal that propagation- and bend- losses are comparable to  that of other antiresonant dielectric waveguides -- but with the added benefit of having immediate exposure to the surrounding environment and direct access to the waveguide core, resulting in effectively diffractionless propagation in free space. Exploiting the direct access to the guided mode, we demonstrated  alignment-free sensing of small samples, but the long diffractionless propagation could in principle also be used for increased sensitivity gas sensing, including at high temperatures.

Much can still be done to reduce losses further,  both by lowering the loss of the cladding materials~\cite{islam2021single} or by tailoring the coupled strand resonances~\cite{li2022interpreting}. We believe this will enable these structures to serve as flexible waveguides for constructing modular terahertz sensors and circuits.

\begin{acknowledgement}
A.T. and M.M. are supported by an Australian Research Council Discovery Early Career Researcher Award (DE200101041, DE210100975). The authors acknowledge the facilities as well as the scientific and technical assistance of the Research and Prototype Foundry Core Research Facility  at the University of Sydney, part of the NSW node of the NCRIS-enabled Australian National Fabrication Facility, where part of this work was performed.
\end{acknowledgement}

\begin{suppinfo}
Animation of the measured and simulated electric fields in Fig.~3(b) and 3(c) (straight waveguides).\\
Animation of the measured and simulated electric fields in Fig.~4(a) and 3(b) (bent waveguides).
\end{suppinfo}

\bibliography{achemso-demo}

\end{document}